\documentclass[%
 reprint,
 showpacs,
 amsmath,amssymb,
 aps,
]{revtex4-1}

\usepackage{color}

\usepackage{graphicx}
\usepackage{dcolumn}
\usepackage{bm}
\usepackage{hyperref}

\usepackage{gensymb}
\usepackage{color}

\usepackage{subcaption}

\begin{document}


\title{Finite Element Lattice Boltzmann Simulations of Contact Line Dynamics}%
\author{Rastin Matin}
 \email{rastin@nbi.ku.dk}
\author{Marek Krzysztof Misztal}%
 \email{misztal@nbi.ku.dk}
\author{Anier Hern\'{a}ndez-Garc\'{i}a}
\author{Joachim Mathiesen}

\affiliation{%
 Niels Bohr Institute, University of Copenhagen, DK-2100 Copenhagen, Denmark
}%

\date{\today}%

\begin{abstract}
The lattice Boltzmann method has become a standard technique for simulating a wide range of fluid flows. However, the intrinsic coupling of momentum and space discretization restricts the traditional lattice Boltzmann method to regular lattices. Alternative off-lattice Boltzmann schemes exist for both single- and multiphase flows that decouple the velocity discretization from the underlying spatial grid. The current study extends the applicability of these off-lattice methods by introducing a finite element formulation that enables simulating contact line dynamics for partially wetting fluids. This work exemplifies the implementation of the scheme and furthermore presents benchmark experiments that show the scheme reduces spurious currents at the liquid-vapor interface by two orders of magnitude compared to a nodal implementation and allows for predicting the equilibrium states accurately in the range of moderate contact angles.

\end{abstract}

\maketitle


\section{\label{sec:level1}Introduction}
The interaction between a liquid and a solid surface plays an important role in many fields ranging in scale from below those considered in microfluidics to scales beyond those in reservoir modelling. In this context the static contact angle defines the intersection between a liquid-vapor interface and a solid surface and specifies the degree of wettability of the surface through Young's equation.

The lattice Boltzmann method (LBM) is a popular method within computational fluid dynamics and several regular-grid based multiphase formulations have emerged within the last 15 years that succesfully describe fluid-solid interactions, thereby enabling the simulation of wetting effects. These formulations fall within different categories such as free-energy (\cite{Briant485, PhysRevE.69.031602, PhysRevE.69.031603, PhysRevE.78.017702, doi:10.1142/S0129183109014710, Lee20108045, Connington2013601}) and interparticle-potential (\cite{PhysRevE.76.066701, schmieschek_harting_2011, PhysRevE.88.013008}) schemes and have been used, for example, in studies of droplet spreading (\emph{e.g.} \cite{PhysRevE.88.013008, Connington2013601, SON201542, PhysRevE.87.013301}) and fluid flow in porous media (\emph{e.g.} \cite{Liu2013, Ghassemi2011135, Yiotis200735, Langaas2001, doi:10.1063/1.3225144, FLD:FLD1972, JHarting}). In particular free-energy based LBMs have become a useful tool for the study of wetting phenomena \cite{MP_Huang}. One of the main advantages of these models over other variants such as interparticle potential (Shan-Chen) is that the surface tension is more easily tuned and the kinematic viscosity ratio and density ratio can be chosen independently \cite{MP_Huang}. Free-energy based LBMs enforce the wetting boundary condition on the wall implicitly in the intermolecular force. On regular grids this is accomplished by specifying the terms in the finite-difference derivatives at the solid boundary \cite{doi:10.1142/S0129183109014710, Lee20108045, Connington2013601, Connington2015453}.

A different class of LBM exists generally known as off-lattice Boltzmann methods, where the spatial and temporal discretizations are decoupled for enhanced geometric flexibility. This class consists of finite volume \cite{FLD:FLD1018, Patil20095262, MISZTAL2015316} and finite element schemes \cite{Lee2001336, Lee2003445, 0295-5075-75-3-434, 10.3389/fphy.2015.00050, Wardle2013230, Matin2017281}. However, previous work on multiphase finite element LBM (FE-LBM) has only regarded the intermolecular force term \cite{Wardle2013230, Matin2017281} and liquid-solid interactions have not been accounted for. In this paper we present an extension of our previous characteristic-based FE-LBM scheme \cite{Matin2017281} by implementing wetting boundary conditions and moving the intermolecular force term to the streaming step. Our formulation is shown to further reduce spurious currents at equilibrium compared to the implementation in \cite{Matin2017281}. To the best of our knowledge, this is the first formulation of wetting boundaries in the framework of off-lattice Boltzmann methods. 

The paper is organized as follows: In Section \ref{sec:num_method} the multiphase model and wall boundary conditions are briefly reviewed. This section also details the FE-LBM on unstructured grids. The scheme is validated in Section \ref{sec:res_discuss} and results are summarized in Section \ref{sec:conclusion}.

\section{\label{sec:num_method}Numerical Method}
We consider the diffuse interface model for incompressible immiscible two-phase flows with large density and kinematic viscosity ratios presented in \cite{Lee20108045, Wardle2013230, Connington2013601}. The main aspects of the model are briefly summarized in the following to provide context.

The composition $C\in [0,1]$ is chosen as the volume fraction of the liquid phase, \emph{i.e.}, $C=1$ for the liquid ($l$) phase and $C=0$ for the vapor ($v$) phase. The time evolution of the diffuse interface is governed by a Cahn-Hilliard equation 
\begin{align}
\partial_t C + \mathbf u \cdot \nabla C = M\nabla^2 \mu, \label{eqn:mp_che}
\end{align}
where $M>0$ is the mobility. The chemical potential $\mu$ in Eq. \eqref{eqn:mp_che} follows from the free-energy functional 
\begin{align}
\Psi_b = \int_V{\bigg(\beta C^2(1-C)^2 + \frac{\kappa}{2}|\nabla C|^2 \bigg)dV}, \label{eqn:FE_func_b}
\end{align}
where $E_0(C) = \beta C^2(1-C)^2$ is the bulk free energy density. Here $\beta$ and $\kappa$ are constants related to the surface tension and interface width. From thermodynamics it follows that
\begin{align}
\mu = \frac{\delta \Psi_b}{\delta C} = 2\beta C(C-1)(2C-1) - \kappa\nabla^2 C \label{eqn:mp_mu}
\end{align}
and the plane interfacial profile in equilibrium follows from setting $\mu=0$,
\begin{align}
C(z) = \frac{1}{2} + \frac{1}{2}\tanh\!\bigg(\frac{2z}{\xi}\bigg).
\end{align}
Here $z$ is the coordinate normal to the interface, $\xi=\sqrt{8\kappa/\beta}$ the interface thickness and $\sigma = \sqrt{2\kappa \beta}/6$ its surface tension. Disregarding body forces, the governing macroscopic equations for the incompressible flow are
\begin{align}
\nabla \cdot \mathbf u                                                     &= 0 \label{eqn:mp_ic} \\
\rho\bigg(\! \partial_t \mathbf u + \mathbf u\cdot \nabla \mathbf u \bigg) &= -\nabla p - C \nabla \mu \label{eqn:mp_ns} \\
&\phantom{{}=}+ \nabla \cdot [\eta (\nabla \mathbf u + (\nabla \mathbf u)^T)] \notag
\end{align}
where $p$ is the hydrodynamic pressure, $\eta$ the dynamic viscosity and $\rho = C\rho_l + (1-C)\rho_v$ the density.

\subsection{Lattice Boltzmann Method}
Recovering the Cahn-Hilliard, pressure evolution and momentum equations in a lattice Boltzmann framework can be accomplished by introducing two particle distribution functions $g_\alpha$ and $h_\alpha$. The distribution function $h_\alpha$ recovers the composition that tracks the interface between the two phases and $g_\alpha$ recovers the hydrodynamic flow fields. The lattice Boltzmann equation for each distribution function is given by \cite{Lee20108045, Wardle2013230}
\begin{align}
\partial_t g_\alpha + e_{\alpha i} \partial_i g_\alpha &=  -\Omega_{g_\alpha} + F_{g_\alpha} \label{eqn:g_mp} \\
\partial_t h_\alpha + e_{\alpha i} \partial_i h_\alpha &=  -\Omega_{h_\alpha} + F_{h_\alpha} \label{eqn:h_mp}
\end{align}
where the intermolecular forcing term $F_{\psi_\alpha}$ and BGK-operator $\Omega_{\psi_\alpha}$ for a given distribution function $\psi \in \{g, h\}$ is
\begin{align}
\Omega_{\psi_\alpha} = &+\frac{1}{\lambda}(\psi_\alpha - \psi_\alpha^{\text{eq}}) \label{eq:omega} \\ 
F_{g_\alpha}         = &+(e_{\alpha i}-u_i)\Big[\partial_i \rho c_s^2 (\Gamma_\alpha(\mathbf u)-\Gamma_\alpha(0)) \label{eq:G}  \\
&+ \mu(\partial_i C) \Gamma_\alpha (\mathbf u)\Big] \notag \\
F_{h_\alpha}         = &+(e_{\alpha i}-u_i)\bigg[\partial_i C - \frac{C}{\rho c_s^2}(\partial_i p - \mu\partial_i C)\bigg]\Gamma_\alpha(\mathbf u) \label{eq:H}\\
& + M(\partial_{k}\partial_{k} \mu) \Gamma_\alpha(\mathbf u) \notag
\end{align}
Here $e_{\alpha i}$ denote the discrete particle velocities in directions $\alpha$ of the velocity lattice and the current work employs the D3Q19 lattice. The relaxation parameter $\lambda$ is proportional to the kinematic viscosity $\nu$, $\nu=\eta/\rho = c_s^2 \lambda = c_s^2\tau\delta t $, where we use the definition $\tau\equiv \lambda/\delta t$. It is taken as the harmonic mean of the bulk relaxation parameters $(\tau_l , \tau_v)$ weighted by $C$ \cite{Lee20108045},
\begin{align}
\frac{1}{\tau} = \frac{C}{\tau_l} + \frac{1-C}{\tau_v}.
\end{align}
The equilibrium distribution functions $g_\alpha^{\text{eq}}$ and $h_\alpha^{\text{eq}}$ of Eqs. \eqref{eqn:g_mp}-\eqref{eqn:h_mp} are
\begin{align}
g_\alpha^{\text{eq}} &= w_\alpha  \bigg[p + \rho c_s^2\bigg(\frac{e_{\alpha i}u_i}{c_s^2} + \frac{(e_{\alpha i}e_{\alpha j}-c_s^2\delta_{ij})u_iu_j}{2c_s^4}\bigg)\bigg] \label{eqn::g_eq} \\
h_\alpha^{\text{eq}} &= w_\alpha C \bigg[1+ \frac{e_{\alpha i}u_i}{c_s^2} + \frac{(e_{\alpha i}e_{\alpha j}-c_s^2\delta_{ij})u_iu_j}{2c_s^4}\bigg] \\
&\equiv \Gamma_\alpha(\mathbf u)C \label{eqn::f_eq}
\end{align}
where $w_\alpha$ are the integral weighting factors of the D3Q19 model. Using the Chapman-Enskog expansion, Eqs. \eqref{eqn:g_mp}-\eqref{eqn:h_mp} can be shown to recover Eqs. \eqref{eqn:mp_che} and \eqref{eqn:mp_ic}-\eqref{eqn:mp_ns} to second-order accuracy at low Mach numbers. The hydrodynamic fields are calculated by taking the zeroth and the first moments of the distribution functions \cite{Lee20108045}
\begin{align}
C        &= \sum_\alpha h_\alpha \\
\rho u_i &= \frac{1}{c_s^2}\sum_\alpha e_{\alpha i} g_\alpha  \label{eq:velocity} \\
p        &= \sum_\alpha g_\alpha \label{eq:pressure}
\end{align}

\subsection{Boundary Conditions}
Incorporating contact angles can be accomplished by adding to the free energy functional $\Psi_b$ a surface term $\Psi_s$ which accounts for the interaction between the liquid-vapor interface and solid surface. Expanded as a power series in the composition calculated at the solid surface, $C_s$, the surface term takes the form \cite{RevModPhys.57.827} 
\begin{align}
\Psi_s  = \int_S{\big(\phi_0 - \phi_1 C_s + \phi_2 C_s^2 - \phi_3 C_s^3\big)dS},
\end{align}
where terms up to cubic order are retained. The constants $\phi_i$ for $i\in \{1,2,3\}$ are given by $\phi_0=\phi_1=0$, $\phi_2=\phi_c/2$ and $\phi_3 = \phi_c/3$, where $\phi_c$ is a constant that recovers the desired contact angle \cite{doi:10.1142/S0129183109014710, Lee20108045, Connington2013601}. In the linear and quadratic approximations the liquid phase at the solid surface is enriched relative to the bulk value by the attraction on wetting surfaces and depleted due to the repulsion on non-wetting surfaces \cite{doi:10.1142/S0129183109014710}. This effect is undesirable in the systems that the current work is directed towards, and the cubic approximation is thus utilized which displays equilibrium densities at the solid surface that are equal to the corresponding bulk values. 

The first boundary condition required for Eq. \eqref{eqn:mp_che} ensures no mass flux normal to a solid boundary due to a chemical potential gradient \cite{Lee20108045, Connington2013601}, 
\begin{align}
\mathbf n\cdot \nabla \mu|_S = 0, \label{eqn:bc1}
\end{align} 
and is satisfied when bounce-back is employed at the solid boundary. The second boundary condition is for $\nabla^2 C$ and can be established by minimizing $\Psi_s$ \cite{doi:10.1142/S0129183109014710},
\begin{align}
\mathbf n \cdot \nabla C|_S = \frac{\phi_c}{\kappa}(C_s-C_s^2), \label{eqn:bc2}
\end{align}
where the equilibrium contact angle $\theta_{\text{eq}}$ follows from Young's equation for a given wetting potential $\Omega_c$ $\cos \theta_{\text{eq}} = -\Omega_c = -\phi_c/\sqrt{2\kappa\beta}$. Alternative formulations of this boundary condition exist such as the geometric formulation presented in \cite{PhysRevE.75.046708}.

Simulating wetting effects with the lattice Boltzmann method thus reduces to implicitly imposing Eq. \eqref{eqn:bc2} at the solid boundary in the relevant terms of the intermolecular forces. On regular grids this is accomplished by specifying the terms in the finite-difference derivatives at the solid boundary, see \emph{e.g.}, \cite{doi:10.1142/S0129183109014710, Lee20108045, Connington2013601, Connington2015453}. In the following we describe a method for enforcing them on an irregular grid.

\subsection{\label{sec:FEM}Finite Element Method}
Equations \eqref{eqn:g_mp}-\eqref{eqn:h_mp} can be solved at any point by streaming along characteristics from $(\mathbf{x},t)$ to $(\mathbf{x}+\delta t\mathbf{e}_\alpha, t+\delta t)$ (where $\delta t$ is the time step), and applying the trapezoid rule to the RHS \cite{Lee2003445, Connington2015453, Matin2017281}. 
\begin{align}
&\psi_\alpha(\mathbf{x}+\delta t\mathbf{e}_\alpha, t+\delta t)-\psi_\alpha(\mathbf{x}, t) \label{eqn::integrated}=\\ &\frac{\delta t}{2} (F_{\psi_\alpha}-\Omega_{\psi_\alpha})|_{(\mathbf{x}, t)}+ \frac{\delta t}{2}(F_{\psi_\alpha}-\Omega_{\psi_\alpha})|_{(\mathbf{x}+\delta t\mathbf{e}_\alpha, t+\delta t)}\notag.
\end{align}
By introducing the new variable $\bar{\psi}_{\alpha}(\mathbf{x},t) = \psi_{\alpha} + \frac{\delta t}{2}\Omega_{\psi_\alpha}$ we can recast Eq. \eqref{eqn::integrated} as
\begin{align}
&\bar{\psi}_\alpha(\mathbf{x}+\delta t\mathbf{e}_\alpha, t+\delta t)-\bar{\psi}_\alpha(\mathbf{x}, t) = \label{eqn::new_variable}\\ 
&-\frac{1}{\tau+0.5}(\bar{\psi}_\alpha-\bar{\psi}_\alpha^{\text{eq}})+\frac{\delta t}{2} \left(F_{{\psi}_\alpha}|_{(\mathbf{x}, t)}+F_{\psi_\alpha}|_{(\mathbf{x}+\delta t\mathbf{e}_\alpha, t+\delta t)}\right)\notag
\end{align}
where $\bar{\psi}^{\text{eq}}_\alpha = \psi^{\text{eq}}_\alpha$, $\tau = \lambda/{\delta t}$, and the moments of $\bar{\psi}$ recover the same macroscopic fields as the moments of $\psi$. The force term can be treated in an implicit manner and integrated locally in the collision step \cite{Matin2017281}. It is observed, however, that integrating the force term in the streaming step as done in \cite{Wardle2013230} greatly enhances stability when simulating surface wettability, and allows to minimize spurious currents through careful selection of the spatial discretization scheme (as discussed in greater detail below).

Equation \eqref{eqn::new_variable} is solved at a grid point $\mathbf{x}_i$ using the standard two-step procedure:\\
\paragraph*{Collision}
\begin{align}
\hat{\psi}_\alpha^n = \bar{\psi}_\alpha^n - \frac{1}{\tau+0.5} \left(\bar{\psi}_\alpha^n-\bar{\psi}_\alpha^{\text{eq},n} \right) \label{eqn::collide}
\end{align}
\paragraph*{Streaming}
\begin{align}
\bar{\psi}_\alpha^{n+1} &= \hat{\psi}_\alpha^n  - \delta t \big(e_{\alpha i}\partial_i \hat{\psi}_\alpha^n - F^{n}_{\psi_\alpha}\big) \label{eqn::stream} \\
     &\phantom{{}=f_\alpha^n} + \frac{\delta t^2}{2} e_{\alpha j}\partial_j\big(e_{\alpha i}\partial_i \hat{\psi}_\alpha^n - F^{n}_{\psi_\alpha}\big) \notag 
\end{align}
where the superscripts correspond to the time step. Equation \eqref{eqn::stream} is obtained from \eqref{eqn::new_variable} by approximating
\begin{align}
\hat{\psi}^n(\mathbf{x}_i-\delta t \mathbf{e}_\alpha) &= \hat{\psi}^n(\mathbf{x}_i)-\delta t e_{\alpha i}\partial_i\hat{\psi}_\alpha^n\\
&\phantom{{}=\hat{\psi}^n(\mathbf{x}_i)}+\frac{\delta t^2}{2}e_{\alpha j}e_{\alpha_i}\partial_j \partial_i\hat{\psi}_\alpha^n+\mathcal{O}(\delta t^3), \notag\\
F^n_{\psi_\alpha}(\mathbf{x}_i-\delta t\mathbf{e}_\alpha) &= F^n_{\psi_\alpha}(\mathbf{x}_i)-\delta t e_{\alpha i}\partial_i F^n_{\psi_\alpha} + \mathcal{O}(\delta t^2),\\
F^{n+1}_{\psi_\alpha}(\mathbf{x}_i) &= F^n_{\psi_\alpha}(\mathbf{x}_i)+\mathcal{O}(\delta t). \label{eqn:force_apprx}
\end{align}
We note that $F^{n+1}_{\psi_\alpha}(\mathbf{x}_i)$ is approximated by $F^n_{\psi_\alpha}(\mathbf{x}_i)$ in order to avoid implicitness. This approximation still yields a second-order accurate and conditionally stable expression in time (see appendix of \cite{Lee_Lin_Chen_2006}). Intuitively, the approximation follows from the observation that the macroscopic fields recovered by $\tilde \psi_\alpha$ change at a much slower rate than the individual populations.

Equation \eqref{eqn::stream} is discretized in space using the Galerkin finite element method, where spatial decomposition using linear, tetrahedral elements has been applied. The particle distribution functions are specified at the mesh nodes (vertices) $\mathbf x_1, \mathbf x_2, \ldots, \mathbf x_{N_V}$ and interpolated at other points,
\begin{align}
\bar{\psi}_\alpha(\mathbf{x}) \approx \tilde{\psi}_\alpha^n(\mathbf{x}) = \mathbf{N}^{(1)}(\mathbf{x})^T \boldsymbol{\psi}_\alpha^n, \label{eq:discrete_field}
\end{align} 
where $\tilde{\psi}_\alpha^n(\mathbf{x})$ is the approximate solution evaluated at a point $\mathbf x$, and $\boldsymbol{\psi}^n_\alpha = \left[\tilde{\psi}_\alpha^n(\mathbf x_{1}), \tilde{\psi}_\alpha^n(\mathbf x_{2}), \ldots, \tilde{\psi}_\alpha^n(\mathbf x_{N_V})\right]^T$ is the vector of the nodal values. Furthermore, $\mathbf{N}^{(1)}(\mathbf{x}) = \left[\phi_1^{(1)}(\mathbf{x}), \phi_2^{(1)}(\mathbf{x}), \ldots, \phi_{N_V}^{(1)}(\mathbf{x})\right]^T$ where $\phi_i^{(1)}$ is the piecewise-linear shape function corresponding to the node $\mathbf{x}_i$, satisfying $\phi_{i}^{(1)}(\mathbf{x}_j)=\delta_{ij}$ by construction.

Special care must be taken when discretizing the force terms in order to reduce spurious currents in the interfacial region and prevent artificial behaviour, such as mass diffusion between phases (and corresponding thickening of the interface). While Wardle and Lee \cite{Wardle2013230} suggest to use the same discretization scheme for the force terms, experiments show that this approach suffers from such unwanted behaviour. An accurate approach can be derived by analysing the numerical method in case of a static interface at equilibrium, in absence of surface tension, i.e. $\mathbf{u}(\mathbf{x}) = 0$, $p(\mathbf{x}) = 0$, $\kappa = 0$. In such case $F_{f_\alpha} = w_\alpha e_{\alpha i} \partial_i C$, $F_{g_\alpha} = 0$. Naturally, the collision and streaming equations for $g_\alpha$ preserve zero pressure field. However, when the concentration gradient is non-zero, the streaming term in the equation for $f_\alpha$ results in the diffusion of concentration from the heavier (liquid) phase ($C = 1$) into the lighter (vapor) phase ($C = 0$). This can be balanced out by choosing the discretization of $\tilde{\nabla} C^n$, such that 
\begin{align}
\sum_{\alpha=0}^{K} \int_{\Omega_k} \mathbf{e}_\alpha \cdot\left(\nabla \mathbf{N}^{(1)}(\mathbf{x})^T \boldsymbol{\psi}_\alpha^n-\tilde{\nabla} C^n\right)d\Omega = 0,
\end{align}
where $\Omega_k$ is the $k$-th element (tetrahedron) in the mesh. In our setting, one such discretization is the piecewise constant one, i.e. $\tilde{\nabla} C^n = \mathbf{N}^{(0)}(\mathbf{x})^T \boldsymbol{\nabla C}^n$, where $\boldsymbol{\nabla C}^n = \left[\nabla C^n_1, \nabla C^n_2, \ldots, \nabla C^n_{N_T} \right]^T$, $\nabla C^n_k$ is the constant concentration gradient inside an element $\Omega_k$, computed from the nodal values of the piecewise linear concentration field $C^n(\mathbf{x}_i) = \sum_{\alpha=0}^{K} \tilde{f}_\alpha^n (\mathbf{x}_i)$, and $\mathbf{N}^{(0)}(\mathbf{x})^T = \left[ \phi_1^{(0)}, \phi_2^{(0)}, \ldots, \phi_{N_T}^{(0)} \right]$, where $\phi_k^{(0)}$ is the piecewise-constant shape function, $\phi_k^{(0)}(\mathbf{x})\big|_{\Omega_l} \equiv \delta_{kl}$.

The above reasoning suggests to use a piecewise linear discretization for the fields defined through algebraic operations on the distribution functions $\psi_\alpha$ (i.e. $C$, $p$, $\mathbf{u}$, $\mu$ and $\rho$) and a piecewise constant discretization for the force terms $F_{\psi_\alpha}$: $F_{\psi_\alpha}(\mathbf{x}) = \mathbf{N}^{(0)}(\mathbf{x})^T \boldsymbol{\Phi}_{\psi_\alpha}^n$, where 
\begin{align}
\boldsymbol{\Phi}_{\psi_\alpha}^n &= \left[ \Phi^n_{\psi_\alpha, 1}, \Phi^n_{\psi_\alpha, 2}, \ldots, \Phi^n_{\psi_\alpha, N_T} \right]^T \\
\Phi^n_{\psi_\alpha, k} &= \frac{1}{|\Omega_k|} \int_{\Omega_k} F_{\psi_\alpha} d\Omega \label{eq:intF}
\end{align} 
The integral in Eq. \eqref{eq:intF} above can be approximated with great accuracy within the expected range of values of the hydrodynamic fields by substituting the barycentric values of the piecewise-linear fields into Eqs. $\eqref{eq:G}$ and $\eqref{eq:H}$. Based on analytical calculations, the largest error within the expected range of physical variables due to this approximation is found to be roughly 1\%.

Finally, the weak form of Eq. \eqref{eqn::stream} reads
\begin{align}
\mathbf{M}^{(1,1)} ({\boldsymbol{\psi}}_\alpha^{n+1} - {\boldsymbol{\psi}}_\alpha^n) = \, &\Big(\!-\delta t \mathbf{C}_\alpha - \delta t^2 \mathbf{D}_\alpha \Big) {\boldsymbol{\psi}}_\alpha^n - \label{eq:galerkin_lbm} \\
&\Big(\!\delta t \mathbf{M}^{(1,0)}  - \delta t^2 \mathbf{K}_\alpha \Big) {\boldsymbol{\Phi}}_{\psi_\alpha}^n  \notag
\end{align}
where matrices $\mathbf{M}^{(1,1)}, \mathbf{C}_\alpha,  \mathbf{D}_\alpha \in \mathbb{R}^{N_V \times N_V}$ and $ \mathbf{M}^{(1,0)}, \mathbf{K}_\alpha \in \mathbb{R}^{N_V \times N_T}$ are defined as
\begin{align}
\mathbf{M}^{(1,1)} &= \int_{\mathcal{D}} \mathbf{N}^{(1)}{\mathbf{N}^{(1)}}^{T} d\Omega \\
\mathbf{M}^{(1,0)} &= \int_{\mathcal{D}} \mathbf{N}^{(1)}{\mathbf{N}^{(0)}}^{T} d\Omega \\
\mathbf{C}_\alpha &= \int_{\mathcal{D}} \mathbf{N}^{(1)} e_{\alpha r} \partial_r {\mathbf{N}^{(1)}}^T d\Omega \\
\mathbf{K}_\alpha &= \frac{1}{2} \int_{\mathcal{D}} \partial_r\mathbf{N}^{(1)} e_{\alpha r} {\mathbf{N}^{(0)}}^T d\Omega \\
\mathbf{D}_\alpha &= \frac{1}{2} \int_{\mathcal{D}} \partial_s\mathbf{N}^{(1)} e_{\alpha s}e_{\alpha r} \partial_r {\mathbf{N}^{(1)}}^T d\Omega
 \label{eq:fem_matrices}
\end{align}
for the domain $\mathcal D$. The linear system is solved using the preconditioned conjugate gradient method, using the lumped mass vector as the preconditioner.

\subsection{\label{subsec:disc}Evaluating Laplacians}
\begin{figure}[h]
\includegraphics[width=0.85\linewidth]{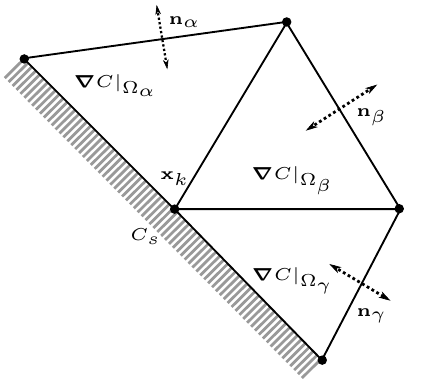}
\caption{\label{fig:vertex_nabla_C} Triangular finite elements centered at vertex $\mathbf x_k$ that form the set $\Omega=\Omega_\alpha\cup\Omega_\beta\cup\Omega_\gamma$ at a solid boundary.}
\end{figure}

The Laplacian terms $\nabla^2C$ and $\nabla^2\mu$ in Eqs. \eqref{eqn:mp_mu} and \eqref{eq:H}, respectively, are stored per element. Looking at the concentration-term first, it is evaluated by considering the volume integral of $\nabla^2C$ over the set $\Omega$, which contains all elements $\Omega_\alpha, \Omega_\beta, \ldots$ that share $\mathbf x_k$ as a common vertex
\begin{align}
\int_\Omega{\nabla^2Cd\Omega} = \int_{\partial\Omega}{\boldsymbol \nabla C\cdot \mathbf n dS},
\end{align}
from which we infer
\begin{align}
\nabla^2 C &\approx \frac{1}{\mathcal V(\Omega)}\int_{d\Omega} {\boldsymbol\nabla C\cdot \mathbf ndS} \\
           &=       \frac{1}{\mathcal V(\Omega)}\sum_{i} \int_{e_i}{\boldsymbol\nabla C\cdot \mathbf ndS}. \label{eqn:lapl_int_mp}
\end{align}
The sum in Eq. \eqref{eqn:lapl_int_mp} runs over all outer edges $e_i$ in $\Omega$. For elements where one or more edges $e_i$ are along the solid boundary, the corresponding terms in the integrand in Eq. \eqref{eqn:lapl_int_mp} are substituted by the value in Eq. \eqref{eqn:bc2}. This is illustrated in Fig. \ref{fig:vertex_nabla_C} for a two-dimensional system.

\begin{figure}[h]
\includegraphics[width=0.85\linewidth]{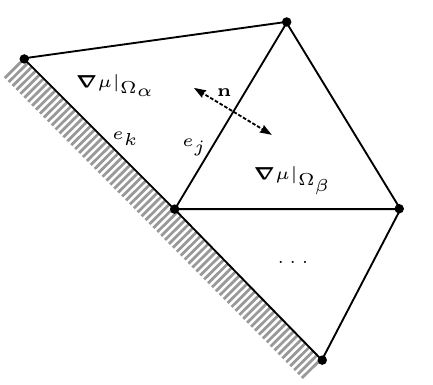}
\caption{\label{fig:element_nabla_mu} Triangular finite elements at a solid boundary. The value at the line segment $e_j$ is taken as $\nabla \mu|_{e_j}=0.5(\boldsymbol\nabla \mu|_{\Omega_\alpha} + \boldsymbol\nabla \mu|_{\Omega_\beta})\cdot \mathbf n$ and $\nabla \mu|_{e_k}=0$.}
\end{figure}

The evaluation of the $\nabla^2\mu$-term follows the same reasoning, but the integration is per element $\Omega_\alpha$
\begin{align}
\nabla^2 \mu &\approx \frac{1}{\mathcal V(\Omega_\alpha)}\int_{d\Omega_\alpha} {\boldsymbol\nabla \mu \cdot \mathbf ndS} \\
             &=       \frac{1}{\mathcal V(\Omega_\alpha)}\sum_{i} \int_{e_i}{\boldsymbol\nabla \mu\cdot \mathbf ndS}
\end{align}
where the integrands along a line segment $e_i$ separating adjacent elements $\Omega_\alpha$ and $\Omega_\beta$ are taken as the average and the value at line segments along the solid boundary are taken as zero as per Eq. \eqref{eqn:bc1}. This is illustrated in Fig. \ref{fig:element_nabla_mu}.

\section{\label{sec:res_discuss}Results and Discussion}
\subsection{Parasitic Currents}
We assess the performance of the scheme by first considering a droplet with radius $R$ in a stationary flow. The average parasitic kinetic energy in the interfacial region $R\pm \xi$ of the droplet is reported in Table \ref{tab:parasitic} for the mixed scheme described in Section \ref{sec:FEM} alongside values obtained using the nodal discretization presented in our earlier work \cite{Matin2017281}. The mixed discretization scheme is observed to succesfully decrease the parasitic currents by two orders of magnitude compared to the nodal discretization. As outlined in Subsection \ref{subsec:disc} this reduction is due to the piecewise constant (linear) discretization of the force terms (physical fields) that balance out the diffusion of concentration from the liquid phase to the vapor phase.
\begin{table}[h!]
\centering
\begin{tabular}{cccc}
  Elements &$\langle \rho \mathbf u\cdot \mathbf u \rangle_N$ &$\langle \rho \mathbf u\cdot \mathbf u \rangle_M$\\
  \hline
  \hline \\ [-2ex]			
	$1.5\cdot 10^6$  &$0.95 \cdot 10^{-6}$  &$3.16\cdot 10^{-9}$ \\
	$2.9\cdot 10^6$  &$0.14 \cdot 10^{-6}$  &$1.61\cdot 10^{-9}$ \\
	$4.9\cdot 10^6$  &$0.048\cdot 10^{-6}$  &$0.65\cdot 10^{-9}$ \\
  \hline
  \hline  
\end{tabular}
\caption{Average parasitic kinetic energy density $\langle \rho \mathbf u\cdot \mathbf u \rangle$ in the region $R\pm \xi$ of a static droplet using the nodal $(N)$ and mixed $(M)$ scheme on different mesh resolutions. The two phases have a density contrast of $\rho_l/\rho_g=2$ and identical kinematic viscosities.}
\label{tab:parasitic}
\end{table}

\subsection{Contact Angle Measurements}
\begin{figure}[h]
\includegraphics[width=0.85\linewidth]{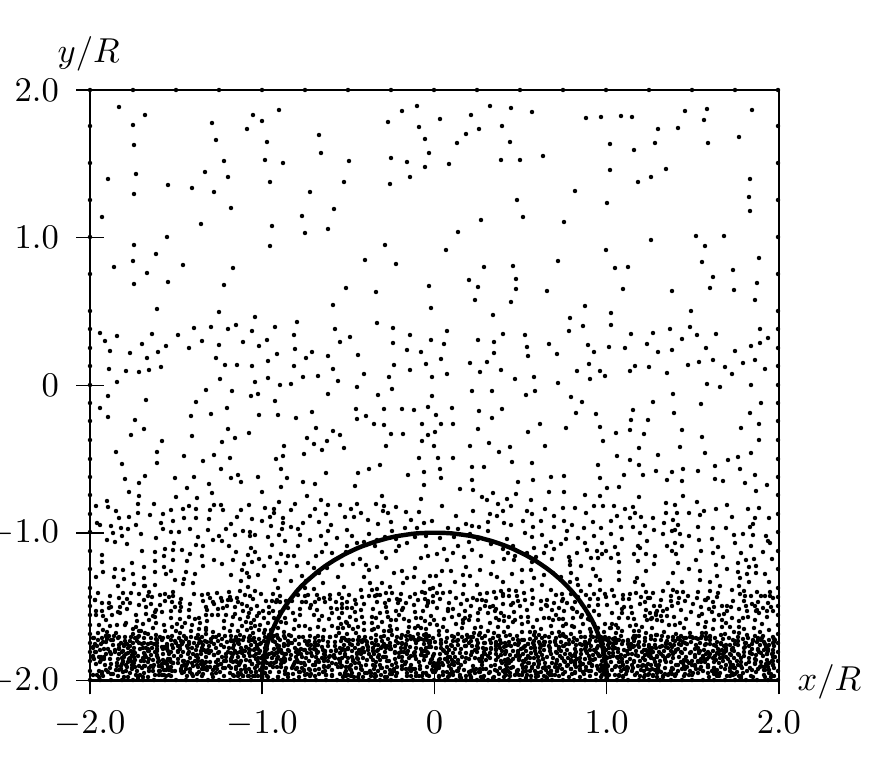}
\caption{\label{fig:mesh} A cross section of the unstructured grid used for the droplet-simulations. The mesh contains $N=1.5\cdot 10^6$ grid points in total. Also shown is the initial contour $C=0.5$.}
\end{figure}

We now turn to an investigation of the equilibrium shape of a three-dimensional droplet on a homogeneous surface. The droplet is initialized as a perfect hemisphere resting on a plane surface with radius $R=1$, see Fig. \ref{fig:mesh}. Also illustrated is the underlying unstructured mesh. It is generated with an increasing resolution towards the bottom surface where the interface dynamics occurs, thereby enhancing the accuracy of the simulation without a significant increase in required computation time.

\begin{figure}[h]
\centering
   \begin{subfigure}[b]{0.4\textwidth}
   \includegraphics[width=0.9\linewidth]{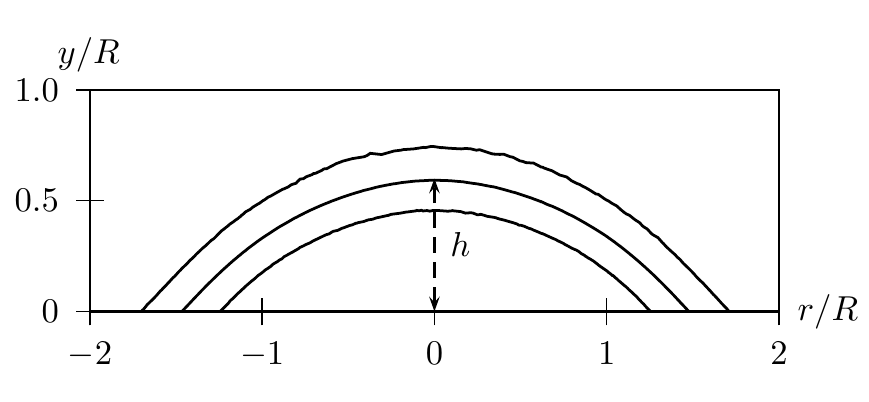}
   \caption{}
   \label{fig:eq_shapes}
\end{subfigure}
\begin{subfigure}[b]{0.4\textwidth}
   \includegraphics[width=0.9\linewidth]{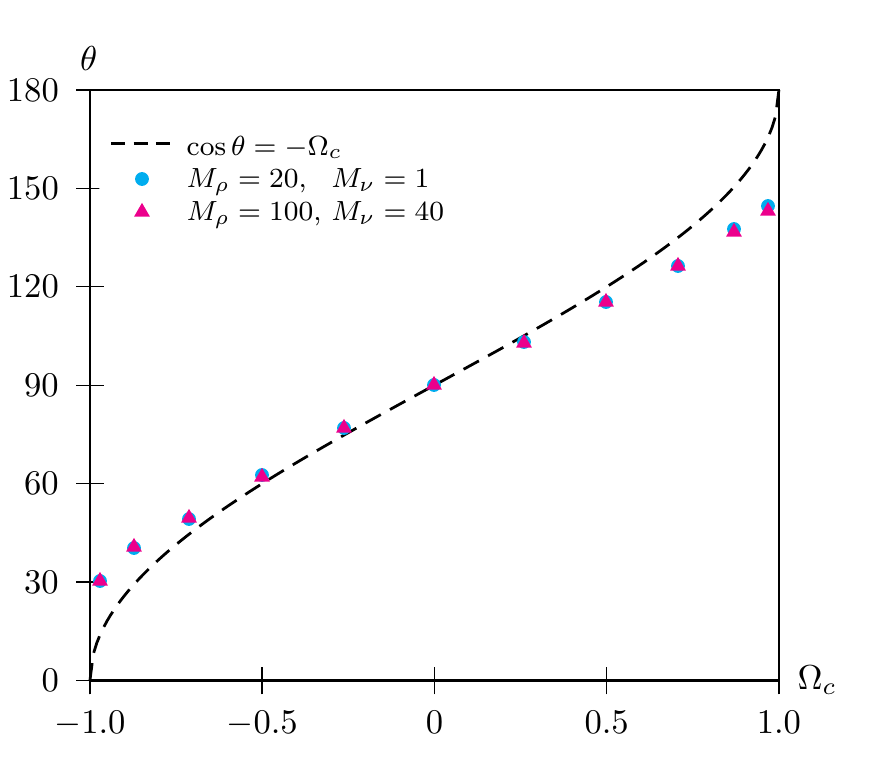}
   \caption{}
   \label{fig:contact_angle} 
\end{subfigure}
\caption{(a) Cross sections of the equilibrium contours ($C=0.1$, $C=0.5$ and $C=0.9$) for a droplet with $\theta_{\text{eq}}=150\degree$ and contrasts $M_\rho = 100$ and $M_\nu=40$. (b) Measured equilibrium contact angle $\theta$ as a function of dimensionless wetting potential $\Omega_c$.}
\label{fig:porous_artificial}
\end{figure}

In Fig. \ref{fig:eq_shapes} three equilibrium contours are shown of a droplet with a wetting potential corresponding to $\theta_{\text{eq}}=150\degree$. It is assumed that the droplet has reached equilibrium when the kinetic energy converges to a steady value asymptotically. The contact angle is then measured from the droplet height $h$ and base diameter $b$ as $\theta=\pi - \arctan(b/2(r-h))$, where $r=(4h^2+b^2)/8h$. Although the interface is several lattice units thick, the measurements are performed on the contour $C=0.5$. 

In Fig. \ref{fig:contact_angle} the full range of angles has been simulated by varying the wetting potential $\Omega_c$ for two set of density- and kinematic viscosity ratios $\{M_\rho, \, M_\nu\}$ and fixed surface tension $\sigma=0.0025$. In general, we obtain good results for moderate contact angles. The largest discrepancy ($23\degree$) appears for a fully non-wet surface. For $45\lesssim \theta_{\text{eq}} \lesssim 135$ the simulated angle is within $5\degree$ of the theoretical value. It has been explicitly verified that these results are independent of the mesh resolution and interface width.

\begin{figure}[h]
\includegraphics[width=0.85\linewidth]{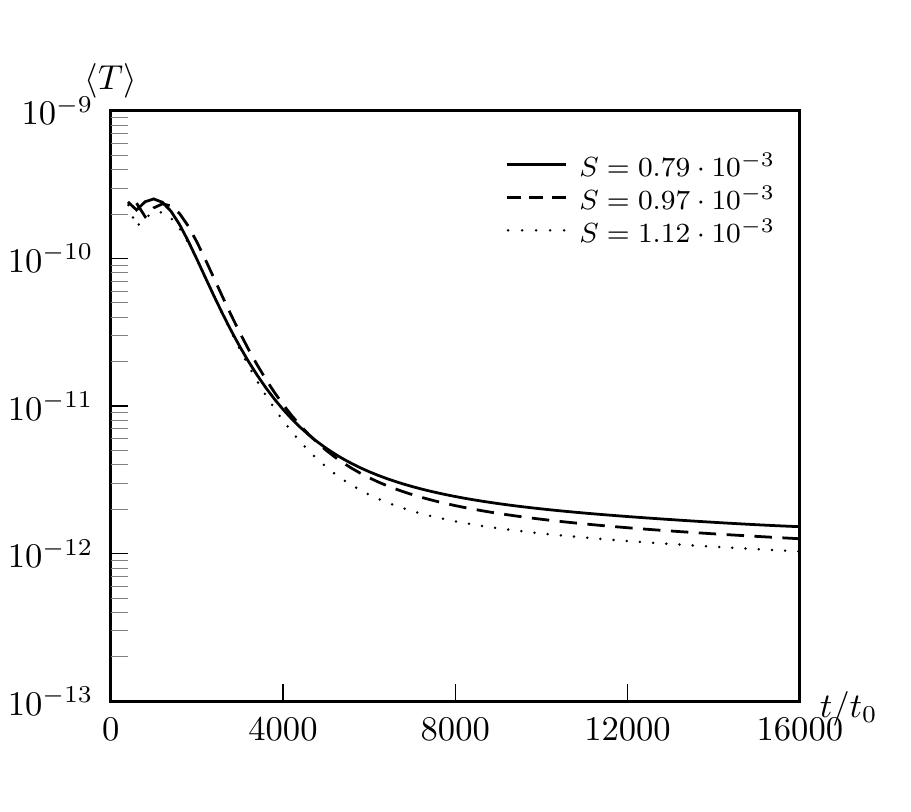}
\caption{\label{fig:mobility}Time evolution of the average kinetic energy density in lattice units $\langle T \rangle = 0.5\sum_i{(\rho_i \mathbf u_i\cdot \mathbf u_i)}/N$ for a droplet with $M_\rho=100$, $M_\nu = 40$ and fixed contact angle $\theta_{\text{eq}}=120\degree$. The mobility $M$ is varied through $S=\sqrt{M\rho_l \nu_l}/R$. Time is scaled by the viscous time $t_0=\rho_l\nu_l R/\sigma$.}
\end{figure}

The time evolution of the kinetic energy per unit volume for $\theta_{\text{eq}}=120\degree$ is shown in Fig. \ref{fig:mobility} for different values of the mobility. Following \cite{yue_zhou_feng_2010, Connington2013601}, the effect of mobility is characterized by the dimensionless number $S=\sqrt{M\rho_l \nu_l}/R$. It is observed that the spurious currents decay at a faster rate for larger values of $S$, \emph{i.e.}, larger values of mobility.

\subsection{Capillary Intrusion}
\begin{figure}[h]
\centering
   \begin{subfigure}[b]{0.4\textwidth}
   \includegraphics[width=0.9\linewidth]{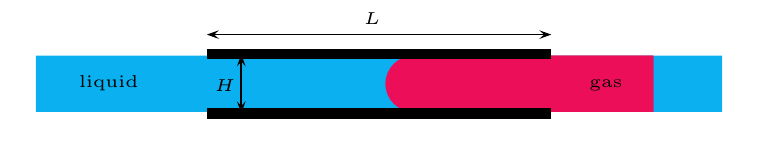}
   \caption{}
   \label{fig:capillary_A} 
\end{subfigure}
\begin{subfigure}[b]{0.4\textwidth}
   \includegraphics[width=0.9\linewidth]{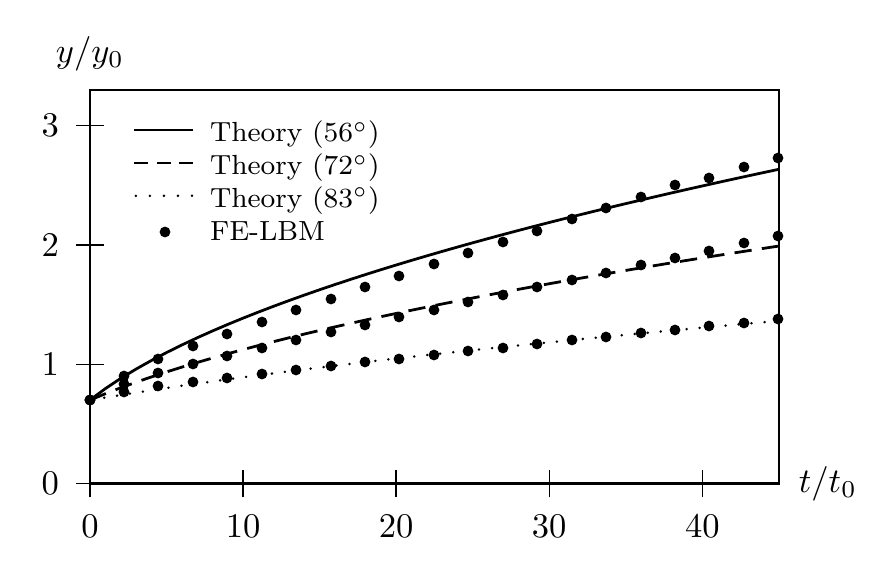}
   \caption{}
   \label{fig:capillary_B}
\end{subfigure}
\caption{(a) Simulation setup for capillary intrusion. (b) The length $y/y_0$ of the column of the intruding liquid as a function of time $t/t_0$. Units are scaled by the characteristic length $y_0=H$ and time $t_0=H\eta_l/\sigma$.}
\label{fig:capillary}
\end{figure}

We now consider the injection of a wetting liquid through a cylindrical capillary tube in order to assess whether the present FE-LBM is able to simulate correct displacement behavior and capture the capillary effect. Following the classical analysis by Washburn \cite{PhysRev.17.273}, we neglect the viscosity of the vapor phase, gravity and intertial effects and furthermore assume that the intruding liquid is incompressible and obeys Poiseuille flow. The average flow velocity of three-dimensional Poiseuille flow is given by
\begin{align}
\langle v \rangle = -\frac{(H/2)^2}{8 \eta_l}\frac{dp}{dx}, \label{eqn:3dpoiseuille}
\end{align}
where $H$ denotes the pipe diameter, $\eta_l$ the liquid dynamic viscosity and $dp/dx$ the pressure gradient that drives the liquid. The Laplace drop across a curved interface is $\Delta p = 2\sigma / R$, where $\sigma$ is the surface tension between the two phases and $R$ the radius of curvature of the interface. The gradient is then $dp/dx = -2\sigma/(Rl)$, where $l$ is the length of the liquid that has penetrated the capillary. The relation between $R$ and the curvature (wetting) angle $\theta$ is $R=H/(2\cos \theta)$. By substituting these relations in Eq. \eqref{eqn:3dpoiseuille} and using $\langle v \rangle = dy/dt$, we obtain the following equation of motion for the interface movement
\begin{align}
\frac{dy}{dt}=\frac{H \sigma \cos \theta}{8\eta_l}.
\end{align}
The simulation setup is illustrated in Fig. \ref{fig:capillary_A}. The mesh contains roughly $10^6$ elements and is periodic along the symmetry axis $(y)$ of the capillary. The middle portion of length $L=20H$ has no-slip wetting boundaries and the boundary conditions are periodic in all directions outside of the middle portion. Fig. \ref{fig:capillary_B} presents the results of our simulations of hydrophilic capillaries with contact angles $\theta=83\degree$, $\theta=72\degree$ and $\theta=56\degree$. The results display good agreement with theory.

\section{\label{sec:conclusion}Conclusion}
In this study a new implementation of the FE-LBM has been developed for simulating liquid droplet behaviour on partial wetting surfaces at large density and kinematic viscosity ratios. The scheme is based on the free-energy multiphase model of Wardle and Lee \cite{Wardle2013230, Lee20108045} and augments our previous nodal-based FE-LBM formulation \cite{Matin2017281} by discretizing the intermolecular forces at elements. Furthermore, the integration of the forces is now performed in the streaming step. 

We have benchmarked our implementation by investigating a liquid droplet in three different settings: Immersed in a vapor phase, resting on a solid surface and moving in a capillary due to capillary pressure. The study reveals that the implementation reduces spurious currents at the interface by two orders of magnitude relative to the nodal implementation in \cite{Matin2017281}. Furthermore, the obtained equilibrium contact angles of the liquid droplet on a solid surface agree within $\pm 5\degree$ with the angles theoretically predicted Young's law for partially wetting fluids ($45\lesssim \theta_{\text{eq}} \lesssim 135$).

In summary, the numerical results indicate that the present FE-LBM scheme is numerically stable and accurate and can be used to study multiphase flows where wetting effects are non-negligble, while harvesting the geometric flexibility of off-lattice schemes. Of particular interest is the effects of reservoir wettability on the relative permeabilities, which is of great importance in reservoir modelling.

\section{Acknowledgments}
The authors acknowledge valuable discussions with Taehun Lee. 
This work is financed by Innovation Fund Denmark and Maersk Oil and Gas A/S through the $\text{P}^3$ project.

\clearpage

\bibliography{unstructured}

\providecommand{\noopsort}[1]{}\providecommand{\singleletter}[1]{#1}%
\begin{thebibliography}{35}%
\makeatletter
\providecommand \@ifxundefined [1]{%
 \@ifx{#1\undefined}
}%
\providecommand \@ifnum [1]{%
 \ifnum #1\expandafter \@firstoftwo
 \else \expandafter \@secondoftwo
 \fi
}%
\providecommand \@ifx [1]{%
 \ifx #1\expandafter \@firstoftwo
 \else \expandafter \@secondoftwo
 \fi
}%
\providecommand \natexlab [1]{#1}%
\providecommand \enquote  [1]{``#1''}%
\providecommand \bibnamefont  [1]{#1}%
\providecommand \bibfnamefont [1]{#1}%
\providecommand \citenamefont [1]{#1}%
\providecommand \href@noop [0]{\@secondoftwo}%
\providecommand \href [0]{\begingroup \@sanitize@url \@href}%
\providecommand \@href[1]{\@@startlink{#1}\@@href}%
\providecommand \@@href[1]{\endgroup#1\@@endlink}%
\providecommand \@sanitize@url [0]{\catcode `\\12\catcode `\$12\catcode
  `\&12\catcode `\#12\catcode `\^12\catcode `\_12\catcode `\%12\relax}%
\providecommand \@@startlink[1]{}%
\providecommand \@@endlink[0]{}%
\providecommand \url  [0]{\begingroup\@sanitize@url \@url }%
\providecommand \@url [1]{\endgroup\@href {#1}{\urlprefix }}%
\providecommand \urlprefix  [0]{URL }%
\providecommand \Eprint [0]{\href }%
\providecommand \doibase [0]{http://dx.doi.org/}%
\providecommand \selectlanguage [0]{\@gobble}%
\providecommand \bibinfo  [0]{\@secondoftwo}%
\providecommand \bibfield  [0]{\@secondoftwo}%
\providecommand \translation [1]{[#1]}%
\providecommand \BibitemOpen [0]{}%
\providecommand \bibitemStop [0]{}%
\providecommand \bibitemNoStop [0]{.\EOS\space}%
\providecommand \EOS [0]{\spacefactor3000\relax}%
\providecommand \BibitemShut  [1]{\csname bibitem#1\endcsname}%
\let\auto@bib@innerbib\@empty
\bibitem [{\citenamefont {Briant}(2002)}]{Briant485}%
  \BibitemOpen
  \bibfield  {author} {\bibinfo {author} {\bibfnamefont {A.}~\bibnamefont
  {Briant}},\ }\href {\doibase 10.1098/rsta.2001.0943} {\bibfield  {journal}
  {\bibinfo  {journal} {Philosophical Transactions of the Royal Society of
  London A: Mathematical, Physical and Engineering Sciences}\ }\textbf
  {\bibinfo {volume} {360}},\ \bibinfo {pages} {485} (\bibinfo {year}
  {2002})}\BibitemShut {NoStop}%
\bibitem [{\citenamefont {Briant}\ \emph {et~al.}(2004)\citenamefont {Briant},
  \citenamefont {Wagner},\ and\ \citenamefont {Yeomans}}]{PhysRevE.69.031602}%
  \BibitemOpen
  \bibfield  {author} {\bibinfo {author} {\bibfnamefont {A.~J.}\ \bibnamefont
  {Briant}}, \bibinfo {author} {\bibfnamefont {A.~J.}\ \bibnamefont {Wagner}},
  \ and\ \bibinfo {author} {\bibfnamefont {J.~M.}\ \bibnamefont {Yeomans}},\
  }\href {\doibase 10.1103/PhysRevE.69.031602} {\bibfield  {journal} {\bibinfo
  {journal} {Phys. Rev. E}\ }\textbf {\bibinfo {volume} {69}},\ \bibinfo
  {pages} {031602} (\bibinfo {year} {2004})}\BibitemShut {NoStop}%
\bibitem [{\citenamefont {Briant}\ and\ \citenamefont
  {Yeomans}(2004)}]{PhysRevE.69.031603}%
  \BibitemOpen
  \bibfield  {author} {\bibinfo {author} {\bibfnamefont {A.~J.}\ \bibnamefont
  {Briant}}\ and\ \bibinfo {author} {\bibfnamefont {J.~M.}\ \bibnamefont
  {Yeomans}},\ }\href {\doibase 10.1103/PhysRevE.69.031603} {\bibfield
  {journal} {\bibinfo  {journal} {Phys. Rev. E}\ }\textbf {\bibinfo {volume}
  {69}},\ \bibinfo {pages} {031603} (\bibinfo {year} {2004})}\BibitemShut
  {NoStop}%
\bibitem [{\citenamefont {Lee}\ and\ \citenamefont
  {Liu}(2008)}]{PhysRevE.78.017702}%
  \BibitemOpen
  \bibfield  {author} {\bibinfo {author} {\bibfnamefont {T.}~\bibnamefont
  {Lee}}\ and\ \bibinfo {author} {\bibfnamefont {L.}~\bibnamefont {Liu}},\
  }\href {\doibase 10.1103/PhysRevE.78.017702} {\bibfield  {journal} {\bibinfo
  {journal} {Phys. Rev. E}\ }\textbf {\bibinfo {volume} {78}},\ \bibinfo
  {pages} {017702} (\bibinfo {year} {2008})}\BibitemShut {NoStop}%
\bibitem [{\citenamefont {Liu}\ and\ \citenamefont
  {Lee}(2009)}]{doi:10.1142/S0129183109014710}%
  \BibitemOpen
  \bibfield  {author} {\bibinfo {author} {\bibfnamefont {L.}~\bibnamefont
  {Liu}}\ and\ \bibinfo {author} {\bibfnamefont {T.}~\bibnamefont {Lee}},\
  }\href {\doibase 10.1142/S0129183109014710} {\bibfield  {journal} {\bibinfo
  {journal} {International Journal of Modern Physics C}\ }\textbf {\bibinfo
  {volume} {20}},\ \bibinfo {pages} {1749} (\bibinfo {year}
  {2009})}\BibitemShut {NoStop}%
\bibitem [{\citenamefont {Lee}\ and\ \citenamefont {Liu}(2010)}]{Lee20108045}%
  \BibitemOpen
  \bibfield  {author} {\bibinfo {author} {\bibfnamefont {T.}~\bibnamefont
  {Lee}}\ and\ \bibinfo {author} {\bibfnamefont {L.}~\bibnamefont {Liu}},\
  }\href {\doibase http://dx.doi.org/10.1016/j.jcp.2010.07.007} {\bibfield
  {journal} {\bibinfo  {journal} {Journal of Computational Physics}\ }\textbf
  {\bibinfo {volume} {229}},\ \bibinfo {pages} {8045 } (\bibinfo {year}
  {2010})}\BibitemShut {NoStop}%
\bibitem [{\citenamefont {Connington}\ and\ \citenamefont
  {Lee}(2013)}]{Connington2013601}%
  \BibitemOpen
  \bibfield  {author} {\bibinfo {author} {\bibfnamefont {K.}~\bibnamefont
  {Connington}}\ and\ \bibinfo {author} {\bibfnamefont {T.}~\bibnamefont
  {Lee}},\ }\href {\doibase http://dx.doi.org/10.1016/j.jcp.2013.05.012}
  {\bibfield  {journal} {\bibinfo  {journal} {Journal of Computational
  Physics}\ }\textbf {\bibinfo {volume} {250}},\ \bibinfo {pages} {601 }
  (\bibinfo {year} {2013})}\BibitemShut {NoStop}%
\bibitem [{\citenamefont {Huang}\ \emph {et~al.}(2007)\citenamefont {Huang},
  \citenamefont {Thorne}, \citenamefont {Schaap},\ and\ \citenamefont
  {Sukop}}]{PhysRevE.76.066701}%
  \BibitemOpen
  \bibfield  {author} {\bibinfo {author} {\bibfnamefont {H.}~\bibnamefont
  {Huang}}, \bibinfo {author} {\bibfnamefont {D.~T.}\ \bibnamefont {Thorne}},
  \bibinfo {author} {\bibfnamefont {M.~G.}\ \bibnamefont {Schaap}}, \ and\
  \bibinfo {author} {\bibfnamefont {M.~C.}\ \bibnamefont {Sukop}},\ }\href
  {\doibase 10.1103/PhysRevE.76.066701} {\bibfield  {journal} {\bibinfo
  {journal} {Phys. Rev. E}\ }\textbf {\bibinfo {volume} {76}},\ \bibinfo
  {pages} {066701} (\bibinfo {year} {2007})}\BibitemShut {NoStop}%
\bibitem [{\citenamefont {Schmieschek}\ and\ \citenamefont
  {Harting}(2011)}]{schmieschek_harting_2011}%
  \BibitemOpen
  \bibfield  {author} {\bibinfo {author} {\bibfnamefont {S.}~\bibnamefont
  {Schmieschek}}\ and\ \bibinfo {author} {\bibfnamefont {J.}~\bibnamefont
  {Harting}},\ }\href {\doibase 10.4208/cicp.201009.271010s} {\bibfield
  {journal} {\bibinfo  {journal} {Communications in Computational Physics}\
  }\textbf {\bibinfo {volume} {9}},\ \bibinfo {pages} {1165–1178} (\bibinfo
  {year} {2011})}\BibitemShut {NoStop}%
\bibitem [{\citenamefont {Jansen}\ \emph {et~al.}(2013)\citenamefont {Jansen},
  \citenamefont {Sotthewes}, \citenamefont {van Swigchem}, \citenamefont
  {Zandvliet},\ and\ \citenamefont {Kooij}}]{PhysRevE.88.013008}%
  \BibitemOpen
  \bibfield  {author} {\bibinfo {author} {\bibfnamefont {H.~P.}\ \bibnamefont
  {Jansen}}, \bibinfo {author} {\bibfnamefont {K.}~\bibnamefont {Sotthewes}},
  \bibinfo {author} {\bibfnamefont {J.}~\bibnamefont {van Swigchem}}, \bibinfo
  {author} {\bibfnamefont {H.~J.~W.}\ \bibnamefont {Zandvliet}}, \ and\
  \bibinfo {author} {\bibfnamefont {E.~S.}\ \bibnamefont {Kooij}},\ }\href
  {\doibase 10.1103/PhysRevE.88.013008} {\bibfield  {journal} {\bibinfo
  {journal} {Phys. Rev. E}\ }\textbf {\bibinfo {volume} {88}},\ \bibinfo
  {pages} {013008} (\bibinfo {year} {2013})}\BibitemShut {NoStop}%
\bibitem [{\citenamefont {Son}\ \emph {et~al.}(2015)\citenamefont {Son},
  \citenamefont {Chen}, \citenamefont {Derome},\ and\ \citenamefont
  {Carmeliet}}]{SON201542}%
  \BibitemOpen
  \bibfield  {author} {\bibinfo {author} {\bibfnamefont {S.}~\bibnamefont
  {Son}}, \bibinfo {author} {\bibfnamefont {L.}~\bibnamefont {Chen}}, \bibinfo
  {author} {\bibfnamefont {D.}~\bibnamefont {Derome}}, \ and\ \bibinfo {author}
  {\bibfnamefont {J.}~\bibnamefont {Carmeliet}},\ }\href {\doibase
  http://dx.doi.org/10.1016/j.compfluid.2015.04.022} {\bibfield  {journal}
  {\bibinfo  {journal} {Computers \& Fluids}\ }\textbf {\bibinfo {volume}
  {117}},\ \bibinfo {pages} {42 } (\bibinfo {year} {2015})}\BibitemShut
  {NoStop}%
\bibitem [{\citenamefont {Wang}\ \emph {et~al.}(2013)\citenamefont {Wang},
  \citenamefont {Huang},\ and\ \citenamefont {Lu}}]{PhysRevE.87.013301}%
  \BibitemOpen
  \bibfield  {author} {\bibinfo {author} {\bibfnamefont {L.}~\bibnamefont
  {Wang}}, \bibinfo {author} {\bibfnamefont {H.-b.}\ \bibnamefont {Huang}}, \
  and\ \bibinfo {author} {\bibfnamefont {X.-Y.}\ \bibnamefont {Lu}},\ }\href
  {\doibase 10.1103/PhysRevE.87.013301} {\bibfield  {journal} {\bibinfo
  {journal} {Phys. Rev. E}\ }\textbf {\bibinfo {volume} {87}},\ \bibinfo
  {pages} {013301} (\bibinfo {year} {2013})}\BibitemShut {NoStop}%
\bibitem [{\citenamefont {Liu}\ \emph {et~al.}(2013)\citenamefont {Liu},
  \citenamefont {Valocchi}, \citenamefont {Kang},\ and\ \citenamefont
  {Werth}}]{Liu2013}%
  \BibitemOpen
  \bibfield  {author} {\bibinfo {author} {\bibfnamefont {H.}~\bibnamefont
  {Liu}}, \bibinfo {author} {\bibfnamefont {A.~J.}\ \bibnamefont {Valocchi}},
  \bibinfo {author} {\bibfnamefont {Q.}~\bibnamefont {Kang}}, \ and\ \bibinfo
  {author} {\bibfnamefont {C.}~\bibnamefont {Werth}},\ }\href@noop {}
  {\bibfield  {journal} {\bibinfo  {journal} {Transport in Porous Media}\
  }\textbf {\bibinfo {volume} {99}},\ \bibinfo {pages} {555} (\bibinfo {year}
  {2013})}\BibitemShut {NoStop}%
\bibitem [{\citenamefont {Ghassemi}\ and\ \citenamefont
  {Pak}(2011)}]{Ghassemi2011135}%
  \BibitemOpen
  \bibfield  {author} {\bibinfo {author} {\bibfnamefont {A.}~\bibnamefont
  {Ghassemi}}\ and\ \bibinfo {author} {\bibfnamefont {A.}~\bibnamefont {Pak}},\
  }\href {\doibase http://dx.doi.org/10.1016/j.petrol.2011.02.007} {\bibfield
  {journal} {\bibinfo  {journal} {Journal of Petroleum Science and
  Engineering}\ }\textbf {\bibinfo {volume} {77}},\ \bibinfo {pages} {135 }
  (\bibinfo {year} {2011})}\BibitemShut {NoStop}%
\bibitem [{\citenamefont {Yiotis}\ \emph {et~al.}(2007)\citenamefont {Yiotis},
  \citenamefont {Psihogios}, \citenamefont {Kainourgiakis}, \citenamefont
  {Papaioannou},\ and\ \citenamefont {Stubos}}]{Yiotis200735}%
  \BibitemOpen
  \bibfield  {author} {\bibinfo {author} {\bibfnamefont {A.~G.}\ \bibnamefont
  {Yiotis}}, \bibinfo {author} {\bibfnamefont {J.}~\bibnamefont {Psihogios}},
  \bibinfo {author} {\bibfnamefont {M.~E.}\ \bibnamefont {Kainourgiakis}},
  \bibinfo {author} {\bibfnamefont {A.}~\bibnamefont {Papaioannou}}, \ and\
  \bibinfo {author} {\bibfnamefont {A.~K.}\ \bibnamefont {Stubos}},\ }\href
  {\doibase http://dx.doi.org/10.1016/j.colsurfa.2006.12.045} {\bibfield
  {journal} {\bibinfo  {journal} {Colloids and Surfaces A: Physicochemical and
  Engineering Aspects}\ }\textbf {\bibinfo {volume} {300}},\ \bibinfo {pages}
  {35 } (\bibinfo {year} {2007})}\BibitemShut {NoStop}%
\bibitem [{\citenamefont {Langaas}\ and\ \citenamefont
  {Papatzacos}(2001)}]{Langaas2001}%
  \BibitemOpen
  \bibfield  {author} {\bibinfo {author} {\bibfnamefont {K.}~\bibnamefont
  {Langaas}}\ and\ \bibinfo {author} {\bibfnamefont {P.}~\bibnamefont
  {Papatzacos}},\ }\href {\doibase 10.1023/A:1012002002804} {\bibfield
  {journal} {\bibinfo  {journal} {Transport in Porous Media}\ }\textbf
  {\bibinfo {volume} {45}},\ \bibinfo {pages} {241} (\bibinfo {year}
  {2001})}\BibitemShut {NoStop}%
\bibitem [{\citenamefont {Huang}\ and\ \citenamefont {yun
  Lu}(2009)}]{doi:10.1063/1.3225144}%
  \BibitemOpen
  \bibfield  {author} {\bibinfo {author} {\bibfnamefont {H.}~\bibnamefont
  {Huang}}\ and\ \bibinfo {author} {\bibfnamefont {X.}~\bibnamefont {yun Lu}},\
  }\href {\doibase 10.1063/1.3225144} {\bibfield  {journal} {\bibinfo
  {journal} {Physics of Fluids}\ }\textbf {\bibinfo {volume} {21}},\ \bibinfo
  {pages} {092104} (\bibinfo {year} {2009})}\BibitemShut {NoStop}%
\bibitem [{\citenamefont {Huang}\ \emph {et~al.}(2009)\citenamefont {Huang},
  \citenamefont {Li}, \citenamefont {Liu},\ and\ \citenamefont
  {Lu}}]{FLD:FLD1972}%
  \BibitemOpen
  \bibfield  {author} {\bibinfo {author} {\bibfnamefont {H.}~\bibnamefont
  {Huang}}, \bibinfo {author} {\bibfnamefont {Z.}~\bibnamefont {Li}}, \bibinfo
  {author} {\bibfnamefont {S.}~\bibnamefont {Liu}}, \ and\ \bibinfo {author}
  {\bibfnamefont {X.-y.}\ \bibnamefont {Lu}},\ }\href {\doibase
  10.1002/fld.1972} {\bibfield  {journal} {\bibinfo  {journal} {International
  Journal for Numerical Methods in Fluids}\ }\textbf {\bibinfo {volume} {61}},\
  \bibinfo {pages} {341} (\bibinfo {year} {2009})}\BibitemShut {NoStop}%
\bibitem [{\citenamefont {Liu}\ \emph {et~al.}(2015)\citenamefont {Liu},
  \citenamefont {Kang}, \citenamefont {Leonardi}, \citenamefont {Schmieschek},
  \citenamefont {Narváez}, \citenamefont {Jones}, \citenamefont {Williams},
  \citenamefont {Valocchi},\ and\ \citenamefont {Harting}}]{JHarting}%
  \BibitemOpen
  \bibfield  {author} {\bibinfo {author} {\bibfnamefont {H.}~\bibnamefont
  {Liu}}, \bibinfo {author} {\bibfnamefont {Q.}~\bibnamefont {Kang}}, \bibinfo
  {author} {\bibfnamefont {C.}~\bibnamefont {Leonardi}}, \bibinfo {author}
  {\bibfnamefont {S.}~\bibnamefont {Schmieschek}}, \bibinfo {author}
  {\bibfnamefont {A.}~\bibnamefont {Narváez}}, \bibinfo {author}
  {\bibfnamefont {B.}~\bibnamefont {Jones}}, \bibinfo {author} {\bibfnamefont
  {J.}~\bibnamefont {Williams}}, \bibinfo {author} {\bibfnamefont
  {A.}~\bibnamefont {Valocchi}}, \ and\ \bibinfo {author} {\bibfnamefont
  {J.}~\bibnamefont {Harting}},\ }\href {\doibase 10.1007/s10596-015-9542-3}
  {\bibfield  {journal} {\bibinfo  {journal} {Computational Geosciences}\ ,\
  \bibinfo {pages} {1}} (\bibinfo {year} {2015})}\BibitemShut {NoStop}%
\bibitem [{\citenamefont {Huang}\ \emph {et~al.}(2015)\citenamefont {Huang},
  \citenamefont {Sukop},\ and\ \citenamefont {Lu}}]{MP_Huang}%
  \BibitemOpen
  \bibfield  {author} {\bibinfo {author} {\bibfnamefont {H.}~\bibnamefont
  {Huang}}, \bibinfo {author} {\bibfnamefont {C.~M.}\ \bibnamefont {Sukop}}, \
  and\ \bibinfo {author} {\bibfnamefont {X.-Y.}\ \bibnamefont {Lu}},\
  }\href@noop {} {\emph {\bibinfo {title} {Multiphase Lattice Boltzmann
  Methods: Theory and Application}}},\ \bibinfo {edition} {1st}\ ed.\ (\bibinfo
   {publisher} {Wiley},\ \bibinfo {year} {2015})\BibitemShut {NoStop}%
\bibitem [{\citenamefont {Connington}\ \emph {et~al.}(2015)\citenamefont
  {Connington}, \citenamefont {Lee},\ and\ \citenamefont
  {Morris}}]{Connington2015453}%
  \BibitemOpen
  \bibfield  {author} {\bibinfo {author} {\bibfnamefont {K.~W.}\ \bibnamefont
  {Connington}}, \bibinfo {author} {\bibfnamefont {T.}~\bibnamefont {Lee}}, \
  and\ \bibinfo {author} {\bibfnamefont {J.~F.}\ \bibnamefont {Morris}},\
  }\href {\doibase http://dx.doi.org/10.1016/j.jcp.2014.11.044} {\bibfield
  {journal} {\bibinfo  {journal} {Journal of Computational Physics}\ }\textbf
  {\bibinfo {volume} {283}},\ \bibinfo {pages} {453 } (\bibinfo {year}
  {2015})}\BibitemShut {NoStop}%
\bibitem [{\citenamefont {Rossi}\ \emph {et~al.}(2005)\citenamefont {Rossi},
  \citenamefont {Ubertini}, \citenamefont {Bella},\ and\ \citenamefont
  {Succi}}]{FLD:FLD1018}%
  \BibitemOpen
  \bibfield  {author} {\bibinfo {author} {\bibfnamefont {N.}~\bibnamefont
  {Rossi}}, \bibinfo {author} {\bibfnamefont {S.}~\bibnamefont {Ubertini}},
  \bibinfo {author} {\bibfnamefont {G.}~\bibnamefont {Bella}}, \ and\ \bibinfo
  {author} {\bibfnamefont {S.}~\bibnamefont {Succi}},\ }\href {\doibase
  10.1002/fld.1018} {\bibfield  {journal} {\bibinfo  {journal} {International
  Journal for Numerical Methods in Fluids}\ }\textbf {\bibinfo {volume} {49}},\
  \bibinfo {pages} {619} (\bibinfo {year} {2005})}\BibitemShut {NoStop}%
\bibitem [{\citenamefont {Patil}\ and\ \citenamefont
  {Lakshmisha}(2009)}]{Patil20095262}%
  \BibitemOpen
  \bibfield  {author} {\bibinfo {author} {\bibfnamefont {D.~V.}\ \bibnamefont
  {Patil}}\ and\ \bibinfo {author} {\bibfnamefont {K.}~\bibnamefont
  {Lakshmisha}},\ }\href {\doibase http://dx.doi.org/10.1016/j.jcp.2009.04.008}
  {\bibfield  {journal} {\bibinfo  {journal} {Journal of Computational
  Physics}\ }\textbf {\bibinfo {volume} {228}},\ \bibinfo {pages} {5262 }
  (\bibinfo {year} {2009})}\BibitemShut {NoStop}%
\bibitem [{\citenamefont {Misztal}\ \emph
  {et~al.}(2015{\natexlab{a}})\citenamefont {Misztal}, \citenamefont
  {Hernandez-Garcia}, \citenamefont {Matin}, \citenamefont {S{\o}rensen},\ and\
  \citenamefont {Mathiesen}}]{MISZTAL2015316}%
  \BibitemOpen
  \bibfield  {author} {\bibinfo {author} {\bibfnamefont {M.~K.}\ \bibnamefont
  {Misztal}}, \bibinfo {author} {\bibfnamefont {A.}~\bibnamefont
  {Hernandez-Garcia}}, \bibinfo {author} {\bibfnamefont {R.}~\bibnamefont
  {Matin}}, \bibinfo {author} {\bibfnamefont {H.~O.}\ \bibnamefont
  {S{\o}rensen}}, \ and\ \bibinfo {author} {\bibfnamefont {J.}~\bibnamefont
  {Mathiesen}},\ }\href {\doibase http://dx.doi.org/10.1016/j.jcp.2015.05.019}
  {\bibfield  {journal} {\bibinfo  {journal} {Journal of Computational
  Physics}\ }\textbf {\bibinfo {volume} {297}},\ \bibinfo {pages} {316 }
  (\bibinfo {year} {2015}{\natexlab{a}})}\BibitemShut {NoStop}%
\bibitem [{\citenamefont {Lee}\ and\ \citenamefont {Lin}(2001)}]{Lee2001336}%
  \BibitemOpen
  \bibfield  {author} {\bibinfo {author} {\bibfnamefont {T.}~\bibnamefont
  {Lee}}\ and\ \bibinfo {author} {\bibfnamefont {C.-L.}\ \bibnamefont {Lin}},\
  }\href {\doibase http://dx.doi.org/10.1006/jcph.2001.6791} {\bibfield
  {journal} {\bibinfo  {journal} {Journal of Computational Physics}\ }\textbf
  {\bibinfo {volume} {171}},\ \bibinfo {pages} {336 } (\bibinfo {year}
  {2001})}\BibitemShut {NoStop}%
\bibitem [{\citenamefont {Lee}\ and\ \citenamefont {Lin}(2003)}]{Lee2003445}%
  \BibitemOpen
  \bibfield  {author} {\bibinfo {author} {\bibfnamefont {T.}~\bibnamefont
  {Lee}}\ and\ \bibinfo {author} {\bibfnamefont {C.-L.}\ \bibnamefont {Lin}},\
  }\href {\doibase http://dx.doi.org/10.1016/S0021-9991(02)00065-7} {\bibfield
  {journal} {\bibinfo  {journal} {Journal of Computational Physics}\ }\textbf
  {\bibinfo {volume} {185}},\ \bibinfo {pages} {445 } (\bibinfo {year}
  {2003})}\BibitemShut {NoStop}%
\bibitem [{\citenamefont {Bardow}\ \emph {et~al.}(2006)\citenamefont {Bardow},
  \citenamefont {Karlin},\ and\ \citenamefont {Gusev}}]{0295-5075-75-3-434}%
  \BibitemOpen
  \bibfield  {author} {\bibinfo {author} {\bibfnamefont {A.}~\bibnamefont
  {Bardow}}, \bibinfo {author} {\bibfnamefont {I.~V.}\ \bibnamefont {Karlin}},
  \ and\ \bibinfo {author} {\bibfnamefont {A.~A.}\ \bibnamefont {Gusev}},\
  }\href {http://stacks.iop.org/0295-5075/75/i=3/a=434} {\bibfield  {journal}
  {\bibinfo  {journal} {Europhysics Letters}\ }\textbf {\bibinfo {volume}
  {75}},\ \bibinfo {pages} {434} (\bibinfo {year} {2006})}\BibitemShut
  {NoStop}%
\bibitem [{\citenamefont {Misztal}\ \emph
  {et~al.}(2015{\natexlab{b}})\citenamefont {Misztal}, \citenamefont
  {Hernandez-Garcia}, \citenamefont {Matin}, \citenamefont {M{\"u}ter},
  \citenamefont {Jha}, \citenamefont {S{\o}rensen},\ and\ \citenamefont
  {Mathiesen}}]{10.3389/fphy.2015.00050}%
  \BibitemOpen
  \bibfield  {author} {\bibinfo {author} {\bibfnamefont {M.~K.}\ \bibnamefont
  {Misztal}}, \bibinfo {author} {\bibfnamefont {A.}~\bibnamefont
  {Hernandez-Garcia}}, \bibinfo {author} {\bibfnamefont {R.}~\bibnamefont
  {Matin}}, \bibinfo {author} {\bibfnamefont {D.}~\bibnamefont {M{\"u}ter}},
  \bibinfo {author} {\bibfnamefont {D.}~\bibnamefont {Jha}}, \bibinfo {author}
  {\bibfnamefont {H.~O.}\ \bibnamefont {S{\o}rensen}}, \ and\ \bibinfo {author}
  {\bibfnamefont {J.}~\bibnamefont {Mathiesen}},\ }\href {\doibase
  10.3389/fphy.2015.00050} {\bibfield  {journal} {\bibinfo  {journal}
  {Frontiers in Physics}\ }\textbf {\bibinfo {volume} {3}} (\bibinfo {year}
  {2015}{\natexlab{b}}),\ 10.3389/fphy.2015.00050}\BibitemShut {NoStop}%
\bibitem [{\citenamefont {Wardle}\ and\ \citenamefont
  {Lee}(2013)}]{Wardle2013230}%
  \BibitemOpen
  \bibfield  {author} {\bibinfo {author} {\bibfnamefont {K.~E.}\ \bibnamefont
  {Wardle}}\ and\ \bibinfo {author} {\bibfnamefont {T.}~\bibnamefont {Lee}},\
  }\href {\doibase http://dx.doi.org/10.1016/j.camwa.2011.09.020} {\bibfield
  {journal} {\bibinfo  {journal} {Computers and Mathematics with Applications}\
  }\textbf {\bibinfo {volume} {65}},\ \bibinfo {pages} {230 } (\bibinfo {year}
  {2013})},\ \bibinfo {note} {special Issue on Mesoscopic Methods in
  Engineering and Science (ICMMES-2010, Edmonton, Canada)}\BibitemShut
  {NoStop}%
\bibitem [{\citenamefont {Matin}\ \emph {et~al.}(2017)\citenamefont {Matin},
  \citenamefont {Misztal}, \citenamefont {Hernandez-Garcia},\ and\
  \citenamefont {Mathiesen}}]{Matin2017281}%
  \BibitemOpen
  \bibfield  {author} {\bibinfo {author} {\bibfnamefont {R.}~\bibnamefont
  {Matin}}, \bibinfo {author} {\bibfnamefont {M.~K.}\ \bibnamefont {Misztal}},
  \bibinfo {author} {\bibfnamefont {A.}~\bibnamefont {Hernandez-Garcia}}, \
  and\ \bibinfo {author} {\bibfnamefont {J.}~\bibnamefont {Mathiesen}},\ }\href
  {\doibase https://doi.org/10.1016/j.camwa.2017.04.027} {\bibfield  {journal}
  {\bibinfo  {journal} {Computers \& Mathematics with Applications}\ }\textbf
  {\bibinfo {volume} {74}},\ \bibinfo {pages} {281 } (\bibinfo {year}
  {2017})}\BibitemShut {NoStop}%
\bibitem [{\citenamefont {de~Gennes}(1985)}]{RevModPhys.57.827}%
  \BibitemOpen
  \bibfield  {author} {\bibinfo {author} {\bibfnamefont {P.~G.}\ \bibnamefont
  {de~Gennes}},\ }\href {\doibase 10.1103/RevModPhys.57.827} {\bibfield
  {journal} {\bibinfo  {journal} {Rev. Mod. Phys.}\ }\textbf {\bibinfo {volume}
  {57}},\ \bibinfo {pages} {827} (\bibinfo {year} {1985})}\BibitemShut
  {NoStop}%
\bibitem [{\citenamefont {Ding}\ and\ \citenamefont
  {Spelt}(2007)}]{PhysRevE.75.046708}%
  \BibitemOpen
  \bibfield  {author} {\bibinfo {author} {\bibfnamefont {H.}~\bibnamefont
  {Ding}}\ and\ \bibinfo {author} {\bibfnamefont {P.~D.~M.}\ \bibnamefont
  {Spelt}},\ }\href {\doibase 10.1103/PhysRevE.75.046708} {\bibfield  {journal}
  {\bibinfo  {journal} {Phys. Rev. E}\ }\textbf {\bibinfo {volume} {75}},\
  \bibinfo {pages} {046708} (\bibinfo {year} {2007})}\BibitemShut {NoStop}%
\bibitem [{\citenamefont {Lee}\ \emph {et~al.}(2006)\citenamefont {Lee},
  \citenamefont {Lin},\ and\ \citenamefont {Chen}}]{Lee_Lin_Chen_2006}%
  \BibitemOpen
  \bibfield  {author} {\bibinfo {author} {\bibfnamefont {T.}~\bibnamefont
  {Lee}}, \bibinfo {author} {\bibfnamefont {C.-L.}\ \bibnamefont {Lin}}, \ and\
  \bibinfo {author} {\bibfnamefont {L.-D.}\ \bibnamefont {Chen}},\ }\href
  {\doibase 10.1016/j.jcp.2005.10.021} {\bibfield  {journal} {\bibinfo
  {journal} {Journal of Computational Physics}\ }\textbf {\bibinfo {volume}
  {215}} (\bibinfo {year} {2006}),\ 10.1016/j.jcp.2005.10.021}\BibitemShut
  {NoStop}%
\bibitem [{\citenamefont {YUE}\ \emph {et~al.}(2010)\citenamefont {YUE},
  \citenamefont {ZHOU},\ and\ \citenamefont {FENG}}]{yue_zhou_feng_2010}%
  \BibitemOpen
  \bibfield  {author} {\bibinfo {author} {\bibfnamefont {P.}~\bibnamefont
  {YUE}}, \bibinfo {author} {\bibfnamefont {C.}~\bibnamefont {ZHOU}}, \ and\
  \bibinfo {author} {\bibfnamefont {J.~J.}\ \bibnamefont {FENG}},\ }\href
  {\doibase 10.1017/S0022112009992679} {\bibfield  {journal} {\bibinfo
  {journal} {Journal of Fluid Mechanics}\ }\textbf {\bibinfo {volume} {645}},\
  \bibinfo {pages} {279–294} (\bibinfo {year} {2010})}\BibitemShut {NoStop}%
\bibitem [{\citenamefont {Washburn}(1921)}]{PhysRev.17.273}%
  \BibitemOpen
  \bibfield  {author} {\bibinfo {author} {\bibfnamefont {E.~W.}\ \bibnamefont
  {Washburn}},\ }\href {\doibase 10.1103/PhysRev.17.273} {\bibfield  {journal}
  {\bibinfo  {journal} {Phys. Rev.}\ }\textbf {\bibinfo {volume} {17}},\
  \bibinfo {pages} {273} (\bibinfo {year} {1921})}\BibitemShut {NoStop}%
\end{thebibliography}%

\end{document}